\documentclass[runningheads]{llncs}
\usepackage{graphicx}
\usepackage{cite}
\usepackage{booktabs} 
\usepackage{amsmath}
\usepackage{multirow}
\usepackage{balance}
\usepackage{stfloats}
\DeclareMathOperator*{\argmax}{arg\,max}

\usepackage{pifont}

\newcommand{\repeatthanks}{\textsuperscript{\thefootnote}}
\begin{document}
\title{Using Image Captions and Multitask Learning for Recommending Query Reformulations}
\titlerunning{Image Captions and Multitask Learning for Query Reformulations}
%
%
\author{Gaurav Verma\inst{1} \and
Vishwa Vinay\inst{1} \and
Sahil Bansal\inst{2}\thanks{denotes equal contribution.} \and
Shashank Oberoi\inst{3}\repeatthanks \and 
Makkunda Sharma\inst{4}\repeatthanks \and 
Prakhar Gupta\inst{5}}
\vspace{-2mm}
%

\authorrunning{Verma, Vinay, Bansal, Oberoi, Sharma, and Gupta}
%
\institute{Adobe Research, India\\
\email{\{gaverma, vinay\}@adobe.com}\\
 \and
IBM Research, India 
\and 
Adobe Inc., India
\and 
Indian Institute of Technology Delhi, India
\and 
Carnegie Mellon University, USA
}
\maketitle              
\begin{abstract}\vspace{-7mm}
Interactive search sessions often contain multiple queries, where the user submits a reformulated version of the previous query in response to the original results. We aim to enhance the query recommendation experience for a commercial image search engine. Our proposed methodology incorporates 
current state-of-the-art practices from relevant literature -- 
the use of generation-based sequence-to-sequence models that capture session context, and a multitask architecture that simultaneously optimizes the ranking of results. We extend this setup by driving the learning of such a model with captions of clicked images as the target, instead of using the subsequent query within the session. Since these captions tend to be linguistically richer, the reformulation mechanism can be seen as assistance to construct more \textit{descriptive} queries. In addition, via the use of a pairwise loss for the secondary ranking task, we show that the generated reformulations are more \textit{diverse}.  
\vspace{-3mm}
\end{abstract}
\section{Introduction}
\vspace{-3mm}
A successful search relies on the engine accurately interpreting the intent behind a user's query and returning likely relevant results ranked high. There has been much progress allowing search engines to respond effectively even to short keyword queries on rare intents \cite{RareQuerySuggestion, ClassificationRareQueries, PersonalizedExpansion}. Despite this, recommendation of queries is an integral part of all search experiences -- either in the form of \textit{query autocomplete} (queries that match the prefix the user has currently typed into the search box) or \textit{query suggestions} (reformulation options once an initial query has been provided). %
In this work, we focus on the query suggestion task.

Original algorithms for this scenario relied on extracting co-occurrence patterns between query pairs, and their constituent terms, within historical logs \cite{jones2006generating, huang2003relevant, fonseca2005concept, beeferman2000agglomerative}. Such methods often work well for frequent queries. Recent work utilizing generative approaches common in natural language processing (NLP) scenarios offer generalization in terms of being able to provide suggestions even for rare queries \cite{mitra2015exploring, cho2014learning}. More specifically, the work by Sordoni et al. \cite{HRED} focuses on generating query suggestions that are aware of the context of the user's current session. The current paper is most similar to this work in terms of motivation and the core technical component.

The experiments described here are based on data from a commercial stock image search engine. In this setting, the items in the index are professionally taken high quality images to be used in commercial publishing material. The users of such a system exhibit similar properties to what might be expected on general purpose search engines - i.e., the use of relatively short queries often with multiple reformulations within a session. 
The logged data therefore contains not only the sequence of within-session queries, but also impression logs listing what images were shown in response to a query and which amongst those were clicked. 

The availability of usage data, which provides implicit relevance signals, allows the building of a query reformulation model that includes aspects that have been shown to be useful in related literature:
session context capturing information from previous queries in the session, as well as properties of relevant results via a multitask component. Building on state-of-the-art models in this manner, we specialize the solution to our setting by utilizing a novel supervision signal for the reformulation model in the form of linguistically rich captions available for the clicked results (in our case, images) across sessions.
%


\begin{figure*}[!t]
    \centering
    \includegraphics[width=0.85\textwidth]{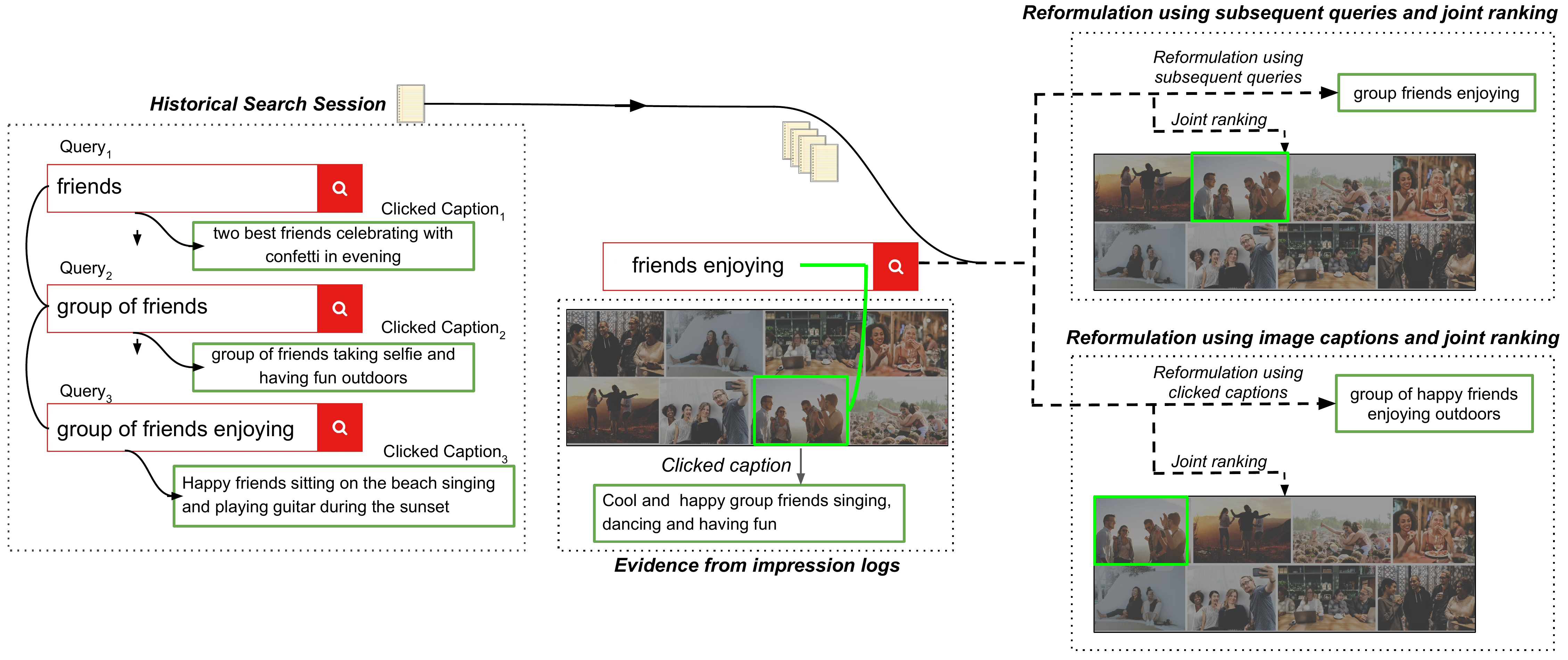}\vspace{-2mm}
\caption{\scriptsize{The basic idea behind our work. We generate query reformulations using \textit{(a)} subsequent queries within sessions, and \textit{(b)} the captions of clicked images, as supervision signals. In both the cases, the task of generating reformulations is done while jointly optimizing the ranking of results.}}
\label{fig:teaser}
\vspace{-5mm}
\end{figure*}

\vspace{-4mm}
\section{Related Work}
\vspace{-3mm}
A user of a search system provides an input query, typically a short list of keywords, into the search box and expects content relevant to their need ranked high in  the result list. There are many reasons why a single iteration of search may not be successful -- mis-specified queries (including spelling errors), imperfect ranking, ambiguous intent, and many more. As a result, it is useful to think of a search session as a series of interactions -- where the user enters a query, examines and potentially interacts with the returned results, and constructs a refined query that is expected to more accurately represent their intent. Search engines therefore mine historical behavior of users on this query and similar ones in an attempt to optimize the entire search session~\cite{silvestri2009mining}. 

Being able to effectively extract these signals from historical logs starts with understanding and interpreting user behavior appropriately. For example, Huang et al. ~\cite{huang2009analyzing} pointed out that successful reformulations, especially those involving changes to words and their order, can be identified as those that retrieve new items which are presented higher in the subsequent results. An automatic reformulation experience involves implementing lessons from such analyses. The first of these is the use of previous queries within the current search sessions to inform the subsequent suggestions – i.e., modeling the \textit{session context}. Earlier papers (e.g.~\cite{cao2008context}) explicitly captured co-occurrence within sessions which, while being an intuitive and simple strategy, had the disadvantage of not being able to account for rarer queries. Newer efforts (e.g.~\cite{mitra2015exploring}) therefore utilize distributed representations of terms and queries to help generalize to unseen queries.

Such efforts are part of a wider expansion of techniques originally common within NLP domains to Information Retrieval (IR) scenarios. 
Conceptually, a generation-based model for query reformulation is obtained by mapping a query to the subsequent one in the same session. Such a model incorporates two signals known to be useful from traditional IR: $(1)$ sequence of terms within a query \& $(2)$ sequence of queries within a session. 
Recent papers have investigated models anchored in the original generic NLP settings but customized to the characteristics of search queries. For example, Dehghani et al. ~\cite{Dehghani2017} suggest a `copy' mechanism within the sequence-to-sequence (seq-to-seq) models \cite{sutskever2014sequence} to allow for terms to be carried over across queries in the session. In the current paper, we consider the work of Sordoni et al. \cite{HRED} as a reference for the core seq-to-seq model. The model, referred to here as  \textit{H}ierarchical \textit{R}ecurrent \textit{E}ncoder \textit{D}ecoder (\textit{\small{HRED}}), is a standard encoder-decoder setup, where word embeddings are aggregated into a query representation, a sequence of which in turn leads to a session representation. A decoder for the hierarchically organized query and session encoders is trained to predict the sequence of query words that compose the subsequent query in the session. Along with being a strong baseline, it serves to illustrate the core components of our work: $(a)$ use of a novel supervision signal in the form of captions of clicked results, and $(b)$ jointly optimizing ranking along with query reformulation. These extensions could similarly be done with other seq-to-seq models used for query suggestion.

Our motivation for using captions of clicked images as supervision signal stems from the fact that captions are often succinct summaries of the content of the actual images as the creators are incentivized to have their images found. In particular, captions indicate which objects are present in the image, their corresponding attributes, as well as relationships with other objects in the same image --  for example, \textit{``A beautiful girl \textbf{wearing} a yellow shirt \textbf{standing near} a red car"}. These properties make the captions a good target.

Multitask learning~\cite{caruana1997multitask} has been shown to have success in scenarios where related tasks benefit from common signals. A recent paper ~\cite{MNSRF} shows benefits of such a pairing in a search setting. Specifically, Ahmad et al. show that coupling with a classifier distinguishing clicked results from those skipped helps improve a query suggestion model. We extend this work by utilizing a pairwise loss function commonly used in learning-to-rank~\cite{Burges2005}. We show that not only does this provide the expected increase in the effectiveness of the ranker component, but also increases the diversity of suggested reformulations. Such diversity has been shown to be important for the query suggestion user experience~\cite{ma2010diversifying}. 

We begin by providing details of the mathematical notation in the next section, before describing our models in detail. The subsequent experimental section provides empirical evidence of the benefits that our design choices bring. 
\vspace{-5mm}
\section{Notation and Model Architectures}
\vspace{-0.5mm}
\label{sec:notations}
\subsubsection{3.1 \quad Notation:}
\vspace{-2mm}
We define a session as a sequence of queries , $\mathcal{S} = \{q_1, \dots, q_n\}$. Each query $q_i$ in session $\mathcal{S}$ has a set of displayed images associated with it, $\mathcal{I}_i = \{I_{i}^1, \dots, I_{i}^m\}$. A subset of images in $\mathcal{I}_i$ are clicked, we refer to the top-ranked clicked image as $I_{i}^{\text{ }\text{clicked}}$. All the images in the set $\mathcal{I}_i$ have a caption describing them, the entire set of which is represented as $\mathcal{C}_i = \{C_{i}^{1}, \dots, C_{i}^{m} \}$. It follows that every $I_i^{\text{ }\text{clicked}}$ will also have an associated caption with it, given as $C_i^{\text{ }\text{clicked}}$. Given this, for every successful query $q_i$ in session $\mathcal{S}$, we will have an associated clicked image $I_i^{\text{ }\text{clicked}}$ and a corresponding caption $C_i^{\text{ }\text{clicked}}$. We consider the size of impression $m$ (number of images) to be fixed for all $q_i$.

Our models treat each query $q_i$ in any given session, as a sequence of words, $q_i = \{w_1, \dots, w_{l_q} \}$. Captions are represented similarly - as sequences of words, $C_i^j = \{w_1, \dots, w_{l_c}\}$.
We use LSTMs \cite{LSTM} to model the sequences, owing to their demonstrated capabilities in modeling various natural language tasks, ranging from machine translation \cite{sutskever2014sequence} to query suggestion \cite{Dehghani2017}.

The input to our models is a query $q_i$ in the session $\mathcal{S}$, and the desired output is a target reformulation $q_{\text{reform}}$. This target reformulation $q_{\text{reform}}$ can either be \textit{(i)} the subsequent query $q_{i+1}$ in the same session $S$, or \textit{(ii)} the caption $C_i^{\text{ } \text{clicked}}$ corresponding to the clicked image $I_i^{\text{ }\text{clicked}}$. Note that obtaining contextual query suggestions via a \textit{translation model} that has learnt a mapping between successive queries within a session (i.e., \textit{(i)}) has been previously proposed in our reference baseline papers~\cite{HRED, MNSRF}. In the current paper, we  utilize a linguistically richer supervision signal, in the form of captions of clicked images (i.e., \textit{(ii)}), and analyze the behavior of the different models across three high level axes - relevance, descriptiveness and diversity of generated reformulations. \vspace{-7mm}
\subsubsection{3.2 \quad Model Architectures: }
\vspace{-2mm}
\label{secModels}
In this paper, we evaluate two base models -- \textit{\small{HRED}} and \textit{\small{HRED}} with \textbf{Cap}tions (\textit{\small{HREDCap}}), and to study the effect of multitask learning, we add a ranker component to each of these models; giving us two more multitask variants -- \textit{\small{HRED + Ranker}} and \textit{\small{HREDCap + Ranker}}. The underlying architecture of \textit{HRED} and \textit{\small{HREDCap}} (and the corresponding variants) is essentially the same, but \textit{\small{HRED}} has been trained by using $q_{i+1}$ as target and \textit{\small{HREDCap}} has been trained using $C_{i}^{clicked}$ as target. \textit{\small{HRED}} comprises of a query encoder, a session encoder, and a query decoder; all of which are descried below.

\begin{figure*}[bt]
    \centering
    \includegraphics[width=1.0\textwidth]{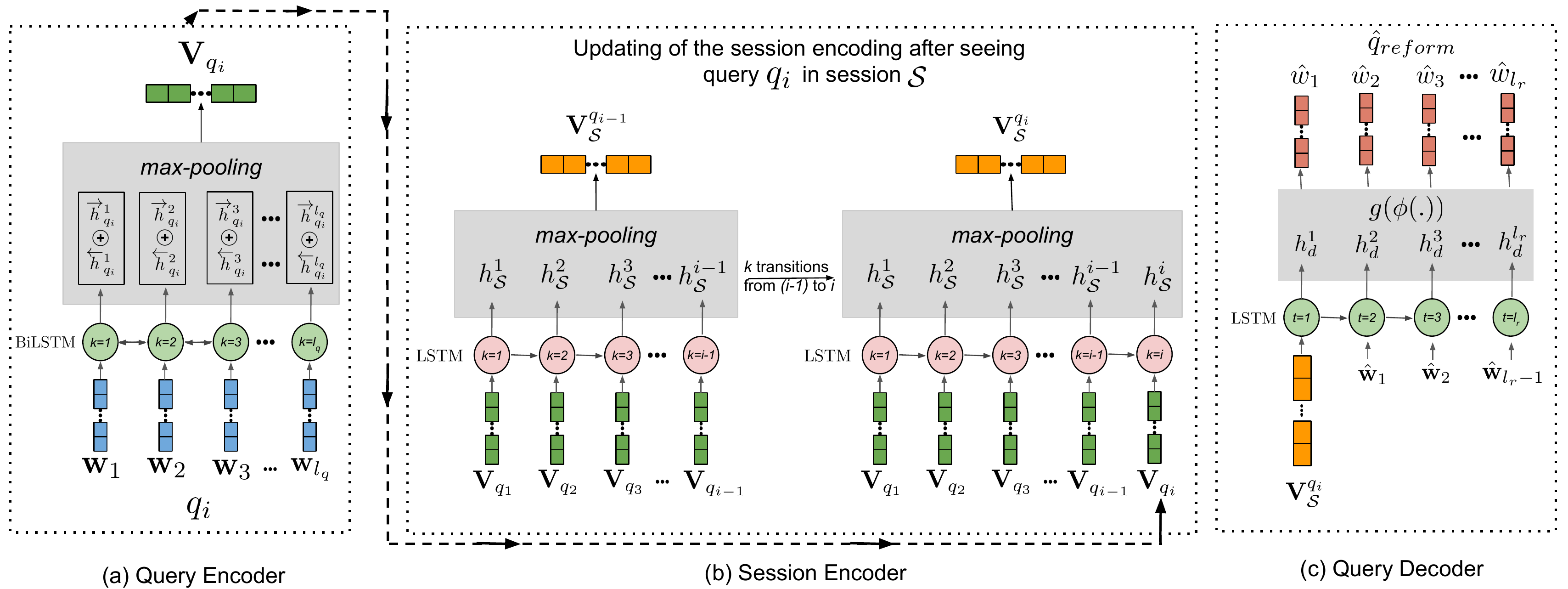}\vspace{-2mm}
    \caption{\scriptsize{An illustration of the \textit{(a)} query encoder,  \textit{(b)} session encoder, and \textit{(c)} query decoder }}
    \label{fig:NotationFigure}
\vspace{-6mm}
\end{figure*}

\noindent\textbf{Query Encoder:} The query encoder generates a query level encoding $\mathbf{V}_{q_i}$ for every $q_i \in \mathcal{S}$. This is done by first representing the query $q_i$ using vector embeddings of corresponding words $\{\mathbf{w}_1, \dots, \mathbf{w}_{l_q}\}$, and then sequentially feeding them into a bidirectional LSTM (BiLSTM) \cite{graves2005framewise}. As shown in Fig. \ref{fig:NotationFigure}(a), the query encoder takes each of these word representations as input to the BiLSTM at every encoding step and updates the hidden states based on the forward and backward pass over the input query. The forward and backward hidden states are concatenated, and after applying attention \cite{bahdanau2014neural} over the concatenated hidden states, we obtain a fixed size vector representation $\mathbf{V}_{q_{i}}$ for the query $q_i \in \mathcal{S}$. 

\noindent\textbf{Session Encoder:} The encoded representation $\mathbf{V}_{q_{i}}$ of query $q_i \in \mathcal{S}$ is used by the session encoder, along with encoded representations  $\{\mathbf{V}_{q_1}, \dots, \mathbf{V}_{q_{i-1}}\}$ of previous queries within the same session, to capture the context of the ongoing session thus far. The session encoder, which is modeled by a unidirectional LSTM \cite{LSTM}, updates the session context $\mathbf{V}^{q_{i}}_{\mathcal{S}}$ after each new $\mathbf{V}_{q_{i}}$ is presented to it. Fig. \ref{fig:NotationFigure}(b) illustrates one such update where the session encoding is updated from $\mathbf{V}^{q_{i-1}}_{\mathcal{S}}$ to $\mathbf{V}^{q_{i}}_{\mathcal{S}}$ after  $\mathbf{V}_{q_{i}}$ is provided as input to the session encoder by the query encoder. 
Since it is unreasonable to assume access to future queries in the session while generating a reformulation for the current query, we use a unidirectional LSTM to model the forward sequence of queries within a session. Accordingly, the session encoder updates its hidden state based on the forward pass over the query sequence. As shown in Fig. \ref{fig:NotationFigure}(b), max-pooling is applied over each dimension of the hidden state to obtain the session encoding  $\mathbf{V}^{q_{i}}_{\mathcal{S}}$. 

\noindent\textbf{Query Decoder:} The generated session encoding $\mathbf{V}^{q_{i}}_{\mathcal{S}}$ is used as input by a query decoder to generate a reformulation $\hat{q}_{\text{reform}} = \{\hat{w}_1, \dots, \hat{w}_{l_r}\}$ for the query $q_i \in \mathcal{S}$. As shown in Fig. \ref{fig:NotationFigure}(c), the reformulation is generated word by word using a single layer unidirectional LSTM. With each unfolding of the decoder LSTM at step $t \in \{1, \dots, l_r\}$, a new word $\hat{w}_t$ is generated as per the following probability:
\vspace{-2mm}\[
    \hat{w}_t =
    \argmax_{w^i \in \mathcal{V}} P(\hat{w}_t = w^i \mid \hat{w}_{1: t-1}, \mathbf{V}^{q_i}_{\mathcal{S}}) \text{ }\text{ } \footnote{For $t=1$, $P(\hat{w}_t = w^i \mid \hat{w}_{1: t-1}, \mathbf{V}^{q_i}_{\mathcal{S}})$ reduces to $P(\hat{w}_t = w^i \mid \mathbf{V}^{q_i}_{\mathcal{S}})$. However, for the sake of readability, this special consideration for $t=1$ has been skipped for the following equations. }
\] \vspace{-2mm}

\vspace{-6mm}\begin{equation}
    P(\hat{w}_t = w^i  \mid \hat{w}_{1: t-1}, \mathbf{V}^{q_i}_{\mathcal{S}}) = g(\phi(h_d^t))
    \label{eq:condGen}\vspace{-1mm}
\end{equation}
Here, $h_d^t$ is the hidden state of the decoder at decoding step $t$, $\hat{w}_{1: t-1}$ denotes the previous words generated by the decoder, and $\phi(h_d^t)$ is a non-linear operation over $h_d^t$.  The softmax function $g(.)$ provides a probability distribution over the entire vocabulary $\mathcal{V}$. $w^i$ is used to denote the $i$-th word in $\mathcal{V}$. The joint probability of generating a reformulation $\hat{q}_{\text{reform}} = \{\hat{w}_1, \dots, \hat{w}_{l_r}\}$ can be decomposed into the ordered conditionals as
    $P(\hat{q}_{\text{reform}} \mid q_i) = \prod_{t = 1}^{l_r} P(\hat{w}_t \mid \hat{w}_{1:t-1}, \mathbf{V}^{q_i}_{\mathcal{S}})$.
During training, the decoder compares each word $\hat{w}_t$ in the generated reformulation $\hat{q}_{\text{reform}}$ with the corresponding word $w_t$ in the target reformulation $q_{\text{reform}}$, and aims to minimize the negative log-likelihood. For a given reformulation by the decoder, the loss is \vspace{-4mm}
\begin{equation}
    \label{eq:reformLoss}
    \mathcal{L}_{\text{reform}} = - \sum_{t = 1}^{l_r} \log P(\hat{w}_t = w_t \mid \hat{w}_{1:t-1}, \mathbf{V}^{q_{i}}_{\mathcal{S}}) + \mathcal{L}_{reg}\vspace{-3mm}
\end{equation}

Here, $\mathcal{L}_{reg} = - \lambda \sum_{w^i \in \mathcal{V}} P(w^i \mid \hat{w}_{1:t-1}, \mathbf{V}^{q_{i}}_{\mathcal{S}}) \cdot \log P(w^i \mid \hat{w}_{1:t-1}, \mathbf{V}^{q_{i}}_{\mathcal{S}})$ is a regularization term added to prevent the predicted probability distribution over the words in the vocabulary from being highly skewed. $\lambda$ is a regularization hyperparameter. The training loss is the sum of $\mathcal{L}_{\text{reform}}$ over all query reformulations generated by the decoder during training.

To summarize, the model encodes the queries, generates session context encodings, and generates the reformulated query using the decoder while updating the model parameters using the gradients of $\mathcal{L}_{\text{reform}}$.

\begin{figure*}[tb]
    \centering
    \begin{minipage}{0.5\textwidth}
        \centering
        \includegraphics[width=1.0\textwidth]{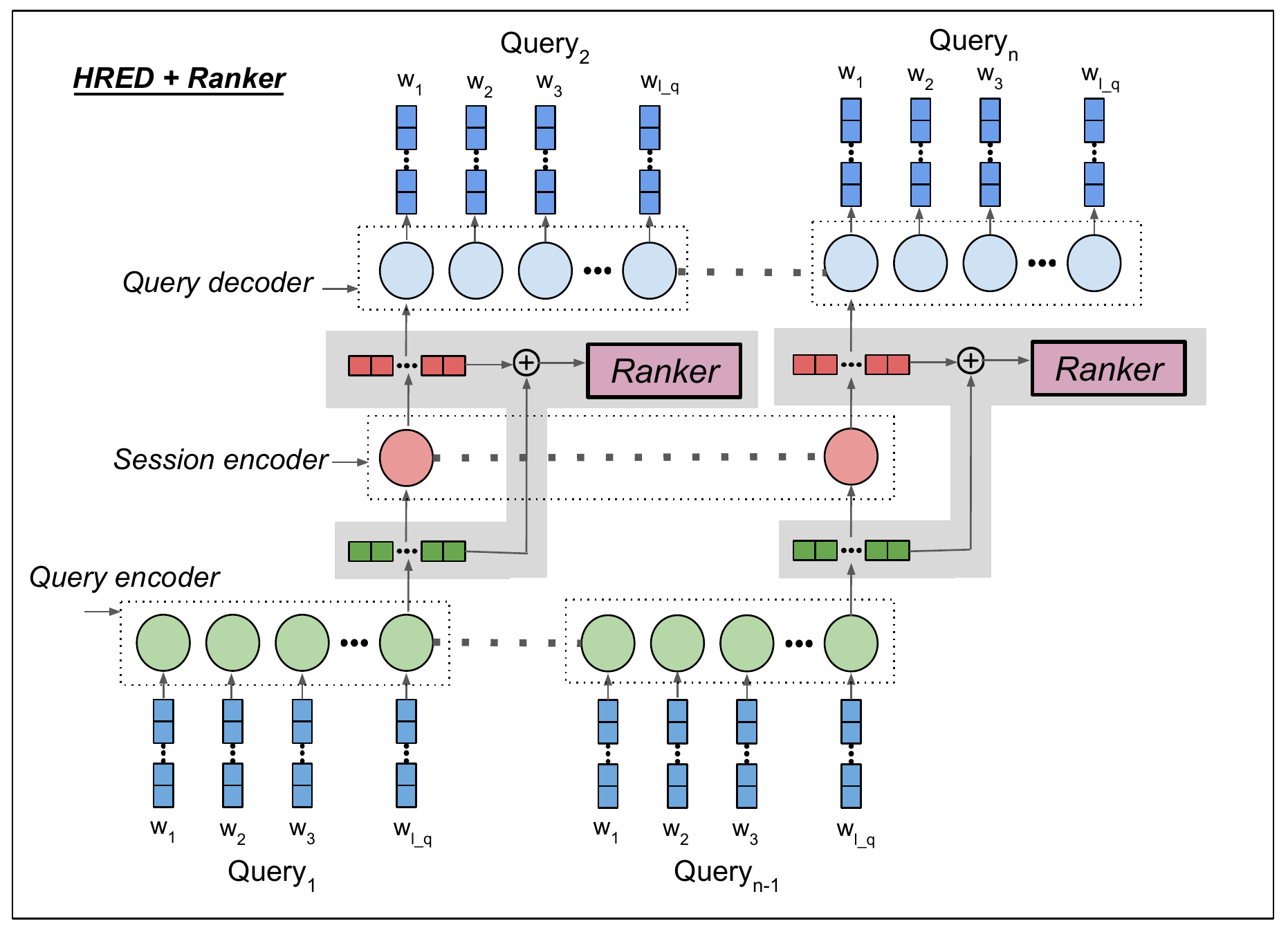}
    \end{minipage}
    \begin{minipage}{0.4\textwidth}
        \centering
        \includegraphics[width=1.0\textwidth]{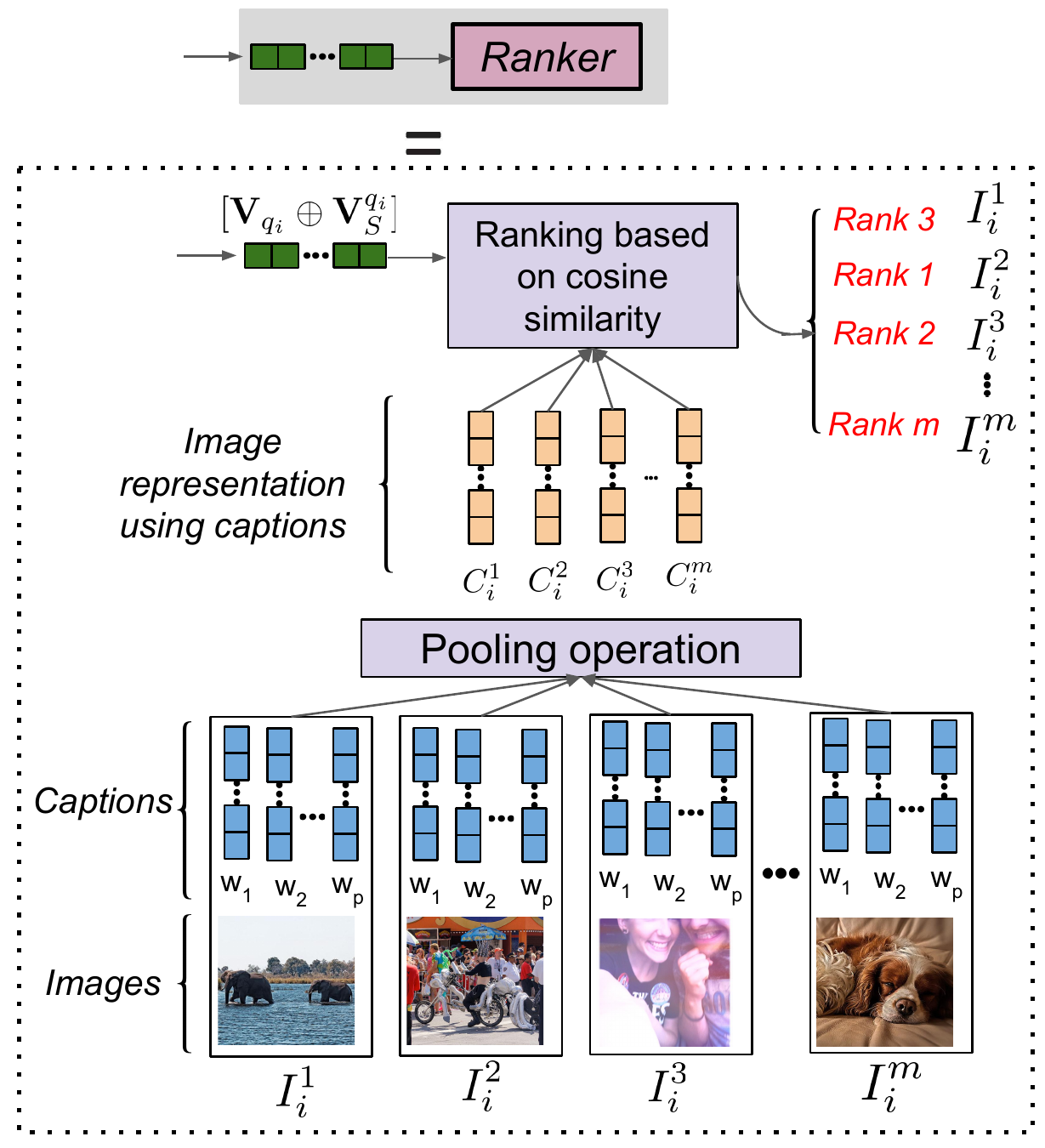}\vspace{-1mm}
    \end{minipage}
\caption{\scriptsize{
The proposed architecture of our multitask model: \textit{HRED + Ranker} (left). For the sake of brevity, we have shown the ranker component separately (right). For \textit{HREDCap + Ranker}, the supervision signals are obtained from captions of clicked images and not subsequent queries.}}
\label{fig:architectures}\vspace{-5mm}
\end{figure*}

\noindent\textbf{Ranker Component:}
This additional component is responsible for ranking the $m$ retrieved results for $q_i \in \mathcal{S}$. As shown in Fig. \ref{fig:architectures} (right), the ranker takes as input the concatenation of query and session encoding $[\mathbf{V}_{q_i} \oplus \mathbf{V}_{\mathcal{S}}^{q_i}]$, for every $q_i \in \mathcal{S}$. 
The concatenated vector representation $[\mathbf{V}_{q_i} \oplus \mathbf{V}_{\mathcal{S}}^{q_i}]$ is used to compute the similarity between the query $q_i$ and its candidate results. The concatenation of these encodings is done to ensure that both current query information (as captured in $\mathbf{V}_{q_i}$) and ongoing session context (as captured in $\mathbf{V}_{\mathcal{S}}^{q_i}$) is used by the ranker.
To obtain a representation of the images, we use their corresponding captions. Formally, for every query $q_i \in \mathcal{S}$ each image $I_i^j \text{ } \in \mathcal{I}_i$ is represented using $\mathbf{C}_i^j$. The average of the vector embeddings of words $\{w_1, \dots, w_{l_c} \}$ in $\mathbf{C}_i^j$ is computed for the image $I_i^j$.  The cosine similarities between $[\mathbf{V}_{q_i} \oplus \mathbf{V}_{\mathcal{S}}^{q_i}]$ and the image representations  $\mathbf{C}_i^j \in \mathcal{C}_i$ are used to rank order the retrieved results. The $j$-th element of the similarity vector $\mathbf{S}_i$ represents the similarity between $[\mathbf{V}_{q_i} \oplus \mathbf{V}_{\mathcal{S}}^{q_i}]$ and $\mathbf{C}_i^j$.
\begin{equation}
\label{similarityEquation}
    {S}_i^j = sim([\mathbf{V}_{q_i} \oplus \mathbf{V}_{\mathcal{S}}^{q_i}], \mathbf{C}_i^j)
\end{equation}

\noindent During training, the ranker tries to learn model parameters based on one of the following two objectives:\\
\noindent(i) \textbf{Cross Entropy Loss}: As described in \cite{MNSRF}, we utilize the `clicked' versus `not-clicked' boolean event to train a classifier, where the ranker scores the $m$ retrieved results based on the probability of being clicked by the user. In the following equation, $\mathbf{R}_i$ for query $q_i$ is an $m$-dimensional vector, where each value in the vector indicates whether the corresponding image was clicked or not. I.e., $R_i^j = 0$ if $I_i^j$ was not clicked, and $R_i^j = 1$ if $I_i^j$ was clicked. A sigmoid of the scores from Eq.~\ref{similarityEquation} is taken as the probability of click. Using the $\mathbf{R}_i$ as labels, the ranker can now be trained using a standard cross entropy loss function:\vspace{-1mm}
\begin{equation}\
\label{eq:rankLoss_BCE}
    \mathcal{L}_{\text{rank}} = BCE(\sigma(\mathbf{S}_i), \mathbf{R}_i)\vspace{-2mm}
\end{equation}

\noindent(ii) \textbf{Pairwise Ranking Loss}: As described in \cite{Burges2005}, the original boolean labels in $\mathbf{R}_i$ can be used to construct an alternate event space where labels $M_{jk} = 1$ when the image at rank $j$ was clicked while the one at $k$ was not. Pairwise ranking loss allows to better model the preferences of certain results over the others.\vspace{-1mm}
\begin{equation}
\label{eq:rankLoss_RO}
    \mathcal{L}_{\text{rank}} = - \frac{1}{m^2} \sum_{j=1}^m\sum_{ \substack{k=1\\k\neq j} }^m M_{jk}*\log \hat{M}_{jk}  + (1 - M_{jk})*\log(1 - \hat{M}_{jk})\vspace{-1mm}
\end{equation}
\[\vspace{-1mm}
\text{where, } \hat{M}_{jk} = P(S_i^j > S_i^k \mid [\mathbf{V}_{q_i} \oplus \mathbf{V}_{\mathcal{S}}^{q_i}]) = \sigma(S_i^j - S_i^k)
\vspace{-1mm}\]

\noindent Since \textit{\small{HRED + Ranker}} and \textit{\small{HREDCap + Ranker}} are multitask models, their training objective is a weighted combination of $\mathcal{L}_{\text{reform}}$ and $\mathcal{L}_{\text{rank}}$. \vspace{-2mm}
\begin{equation}
    \mathcal{L}_{\text{multitask}} = \alpha \cdot \mathcal{L}_{\text{reform}} + (1 - \alpha) \cdot \mathcal{L}_{\text{rank}}\vspace{-1.5mm}
\end{equation}
Here, $\alpha$ is a hyperparameter used for controlling the relative contribution of the two losses. 
As mentioned earlier, either the regular binary cross-entropy loss or the pairwise-ranking loss can be used for $\mathcal{L}_{\text{rank}}$. We experiment using both and report our results on the effect of using one over the other. The models that are trained using cross entropy loss are appended with \textit{(\small{CE})}, and the models that are trained using pairwise ranking objective are denoted as \textit{(\small{RO})}. 

It is worth noting that since for a given query $q_i$ there can be more than one clicked images, our ranker component allows $\mathbf{R}_i$ to take the value $1$ at more than a single place. However, while training the reformulation model, we only consider the caption of the \textit{highest ranked} clicked image. 

%
\vspace{-4mm}
\section{Experiments}
\vspace{-3mm}
\label{sec:experimentalSetup}
\subsubsection{Dataset:} We use logged impression data from Adobe Stock\footnote{\url{https://stock.adobe.com/}}. %
The query logs contain information about the queries that were issued by users, and the images that were presented in response to those queries. Additionally, they contain information about which of the displayed images were clicked by the user. 
We consider the top-10 ranked results, i.e., the number of results to be considered for each query is $m=10$. The queries are segmented into sessions (multiple queries by the same user within a $30$ minute time window), while maintaining the sequence in which they were executed by a user. 
We retain both multi-query sessions as well as single-query sessions, leading to a dataset comprising $1,301,888 $ sessions, $2,122,079$ queries, and $10,185,979$ unique images. We note that $\sim24.8\%$ of the sessions are single-query sessions, while rest all are multi-query sessions; each of which, on average, comprise of $2.19$ queries.
Additionally, we remove all non-alphanumeric characters from the user-entered queries, while keeping spaces, and convert all characters to lowercase.

To obtain the train, test and validation set, we first shuffle the sessions and split them in a $80:10:10$ ratio, respectively. While it is possible for a query to be issued by different users in distinct sessions, a given search session occurs in only one of these sets. These sets are kept the same for all experiments, to ensure consistency while comparing the performance of trained models. The validation set is used for hyperparameter tuning. \vspace{-6.0mm}
\subsubsection{Experimental Setup: }
We construct a global vocabulary $\mathcal{V}$ of size $37,648$ comprising of words that make up the queries and captions for images. Each word in the vocabulary is represented using a $300$-dimensional vector $\mathbf{w}_i$. Each $\mathbf{w}_i \in \mathcal{V}$ is initialized using pre-trained GloVe vectors \cite{pennington2014glove}. Words in our vocabulary $\mathcal{V}$ that do not have a pre-trained embedding available in GloVe ($1,941$ in number), are initialized using samples from a standard normal distribution. 
Since the average number of words in a query, average number of words in a caption, and average number of queries within a session are $2.31$, $5.22$, and $1.63$, we limit their maximum sizes to $5$, $10$, and $5$, respectively. For queries and captions that contain less than $5$ and $10$ words respectively, we pad them using `$<p>$' tokens.  The number of generated words in $\hat{q}_{\text{reform}}$ was limited to $10$, i.e., $l_r = 10$.

During training, we use Adam optimizer~\cite{ADAM} with a learning rate initialized to $10^{-3}$. Across all the models, the regularization coefficient $\lambda$ is set to be $0.1$. For multitask models, the loss trade-off hyperparameter $\alpha$ is set to $0.45$. The sizes of the hidden states of query level encoder $\overrightarrow{h}_q$ and $\overleftarrow{h}_q$ are set to $256$, and that of session level encoder $h_{\mathcal{S}}$ is set to $512$. The size of the decoder's hidden state is kept to be $256$. We train all the models for a maximum of $30$ epochs, using batches of size $512$, with early stopping based on the loss over the validation set. The best trained models are quantitatively and qualitatively evaluated and we discuss the results in the upcoming section. 

At test time, we use 
a beam search-based decoding approach to generate multiple reformulations \cite{bahdanau2014neural}. 
For our experiments, we set the beam width $K=3$. The choice of $K$ was governed by observations that will be discussed later, while analyzing the diversity and relevance of generated reformulations. These three reformulations are rank ordered using their generation probability. 

We experiment with a range of hyperparameters and find that the evaluation results are stable with respect to our hyperparameter choices. However, our motivation is less about training the most accurate models, as we wish to measure the effect of the supervision signal and training objective when used alongside the baseline models. While presenting the results in Table \ref{tab:mainResults} \& \ref{tab:lengtAnalysis}, we report the average of values over $10$ different runs, as well the standard deviations.
\vspace{-5mm}
\section{Evaluation and Results}
\vspace{-3mm}
In this section, we evaluate the performance of the aforementioned models using multiple metrics for each of the two tasks: query reformulation and ranking. The metrics used here are largely inspired from~\cite{Dehghani2017}, and we discuss these below briefly. Towards the end of the section we also provide some qualitative results. \vspace{-3mm}
\vspace{-2mm}
\subsubsection{5.1 \quad Evaluation Metrics:}
Evaluation for query reformulation involves comparing the generated reformulation $\hat{q}_{\text{reform}}$ with the target reformulation $q_{\text{reform}}$. For all the models, irrespective of whether they utilize the next query within the session $q_{i+1}$ as the target reformulation, or the caption $C_i^{\text{ }\text{clicked}}$ corresponding to the clicked image, the ground truth reformulation $q_{\text{reform}}$ is always taken to be $q_{i+1}$\footnote{For sessions with less than $5$ queries in a session, if $q_i$  is the last query of the session, the model is trained to predict the `end of session' token as the first token of $q_{i+1}$. The subsequent predicted tokens are encouraged to be the padding token `$<p>$'.}. This consistency has been maintained across all models to ensure that their performance is comparable, no matter what signal was used to train the reformulation model. The metrics used here cover three aspects: `Relevance' (BLEU \& sim$_{emb}$), `Ranking' (MRR) and `Diversity' (analyzed later).


\noindent\textbf{BLEU score}: This metric~\cite{papineni2002bleu}, commonly used in machine translation scenarios, quantifies the similarity between a predicted sequence of words and the target sequence of words using n-gram precision.
A higher BLEU score corresponds to a higher similarity between the predicted and target reformulations.

\noindent\textbf{Embedding based Query Similarity}: This metric takes semantic similarity of words into account, instead of their exact overlap.
A phrase-level embedding is calculated using vector extrema~\cite{vectorExtrema}, for which pretrained GLoVe embeddings were used. The cosine similarity between the phrase-level vectors for the two queries is given by sim$_{emb}$. A higher value of sim$_{emb}$ is taken to signify a greater semantic similarity between the prediction and the ground truth. Unlike BLEU, we expect sim$_{emb}$ to provide a notion of similarity of the generated query to the target that allows for replacement words that are similar to the observed ones.

\noindent\textbf{Mean Reciprocal Rank (MRR)}: The ranker's effectiveness is evaluated using MRR \cite{MRR}, which is given as the reciprocal rank of the first relevant (i.e., clicked) result averaged over all queries, across all sessions. 
A higher value of MRR will signify a better ranker in the proposed multitask models. To have a standard point of reference to compare against, we computed the observed MRR for the queries in the test set and found it to be $0.31$. This means that on average, for queries in our test set, the first image clicked by the users was at rank $\sim3.1$. \vspace{-3mm}
\vspace{-3mm}

\begin{table*}[!t]
  \centering
  \scalebox{0.71}{
  \begin{tabular}{| c  | c | c | c | c |}\hline
    {} & \multicolumn{3}{c}{\textbf{Query Reformulation}} & \textbf{Ranking}\\
    \textbf{Model}   &  {\textbf{BLEU (\%)} } & {\textbf{sim$_{\mathbf{emb}}$ (\%)} } &\textbf{Diversity}  & \textbf{MRR} \\
    {}   & {($\uparrow$)}  & {($\uparrow$)} &  \textit{Top K = $3$} ($\uparrow$) & \textit{Baseline}: $0.31$ ($\uparrow$)\\\hline
    {HRED}          &  $6.92 \pm 0.06$ & $40.7 \pm 1.3$  & $0.37 \pm 0.01$ & - \\\hline
    {HRED + Ranker} (CE)   &  $7.63 \pm 0.07$ & $43.5 \pm 1.2$  & $0.42 \pm 0.02$ & $0.35 \pm 0.02$ \\
    {HRED + Ranker} (RO)    &  $7.51 \pm 0.07$ & $40.8 \pm 1.4$      & $0.43 \pm 0.02$ & $0.39 \pm 0.01$ \\\hline
    {HREDCap}         &  $7.13 \pm 0.09$ & $37.8 \pm 1.4$      & $0.39 \pm 0.04$ & - \\\hline
    {HREDCap + Ranker} (CE)   &  $7.95 \pm 0.11$ & $39.4 \pm 1.2$      & $0.44 \pm 0.06$ & $0.38 \pm 0.02$ \\
    {HREDCap + Ranker} (RO)   &  $7.68 \pm 0.10$ & $37.6 \pm 1.4$      & $0.45 \pm 0.05$ & $0.41 \pm 0.02$ \\\hline
  \end{tabular}
  }\vspace{0.5mm}
    \caption{\scriptsize{Performance of models based on reformulation and ranking metrics}}\vspace{-9mm}
  \label{tab:mainResults}
\end{table*}

\subsubsection{5.2 \quad Main Results}
Having discussed the metrics, we will now present the performance of our models on the two tasks under consideration, namely query reformulation and ranking. Table \ref{tab:mainResults} provides these results as well as the effect of different ranking losses -- denoted by (\textit{\small{RO}}) and (\textit{\small{CE}}) respectively.

\noindent\textbf{Evaluation based on Reformulation:}
For the purpose of this evaluation, we fix the beam width $K=3$ and report the average of maximum values among all the candidate reformulations, across all queries in our test set. 

While comparing \textit{\small{HRED}} and \textit{\small{HRED + Ranker}} (both \textit{\small{CE}} and \textit{\small{RO}}), we observe that the multitask version performs better across \textit{all} metrics. A similar trend can be observed when comparing \textit{\small{HREDCap}} with its multitask variants. For all the three metrics for query reformulations, the best performing model is a multitask model -- this validates the observations from \cite{MNSRF} in our context. 

When comparing the two core reformulation models -- \textit{\small{HRED}} \& \textit{\small{HREDCap}}, we find that the richer captions data that \textit{\small{HREDCap}} sees is aiding the model -- while \textit{\small{HRED}} scores better sim$_{emb}$, \textit{\small{HREDCap}} wins out on BLEU \& Diversity. The drop in sim$_{emb}$ values can be  explained by noting that on average captions contain more words than queries ($5.22$ in comparison to $2.31$), and hence similarity-based measures, due to additional words in the captions, will not be as high as overlap-based measures (i.e., BLEU).  
\noindent\textbf{Evaluation based on Ranking:}
To evaluate the performance of the ranker component in our proposed multitask models, we use MRR. We use the observed MRR of clicked results in the test set ($0.31$) as the baseline. We also analyze the effect of using the pairwise objective as opposed to the binary cross entropy loss.

Looking at the results presented in Table~\ref{tab:mainResults}, three trends emerge. Firstly, all the proposed multitask models perform better than the baseline. The best performing model, i.e., \textit{\small{HREDCap + Ranker}} with pairwise loss (\textit{\small{RO}}), outperforms the baseline by about $32\%$. Secondly, we observe that using pairwise loss leads to an increase in MRR, for both of the cases under consideration, with only marginal drop in reformulation metrics -- we revisit this observation in the next section. %
Lastly, the multitask models that use captions perform better than multitask models that use subsequent queries. 
\vspace{-3mm}
\subsubsection{5.3 \quad Analysis: }
\vspace{-2mm}
In this section, we concentrate on the following two aspects of the generated query reformulations: $(a)$ diversity, and $(b)$ descriptiveness.

\noindent\textbf{Diverse Query Reformulations due to Multitasking:}
The importance of suggesting diverse queries to enhance user search experience is well established within the IR community. The mechanism to obtain a diverse set of reformulation alternatives is via the use of beam search based decoding.
In scenarios where a set of top-K candidates are required, we take inspiration from Ma et al. \cite{ma2010diversifying} to evaluate the predictions of our models for their diversity. For a beam width of $K$, a reformulation model will generate $\mathcal{R}_{gen}  = \{r_1, r_2, \dots, r_K\}$ candidate reformulations for a given original query. We quantify the diversity in the candidate reformulations by comparing each candidate reformulation $r_i$ with other reformulations $ r_j \in \mathcal{R}_{gen} : i \neq j$. The diversity of a set of $K$ queries is evaluated as 
\vspace{-2mm}\[
    D(\mathcal{R}_{\text{gen}}) = 1 - \frac{1}{K(K-1)}*\left(\sum_{r_i \in \mathcal{R}_{\text{gen}}}\sum_{\substack{r_j \in \mathcal{R}_{\text{gen}}: \text{ }j \neq i}} sim_{emb}(r_i, r_j)\right)\vspace{-2mm}
\]
In Table \ref{tab:mainResults}, it can be observed that multitask models generate more diverse reformulations than models trained just for the task of query reformulation. This is particularly evident when comparing the effect of the ranking loss. 

\begin{figure*}[!h]
\vspace{-6mm}
    \centering
    \includegraphics[width=0.9\textwidth]{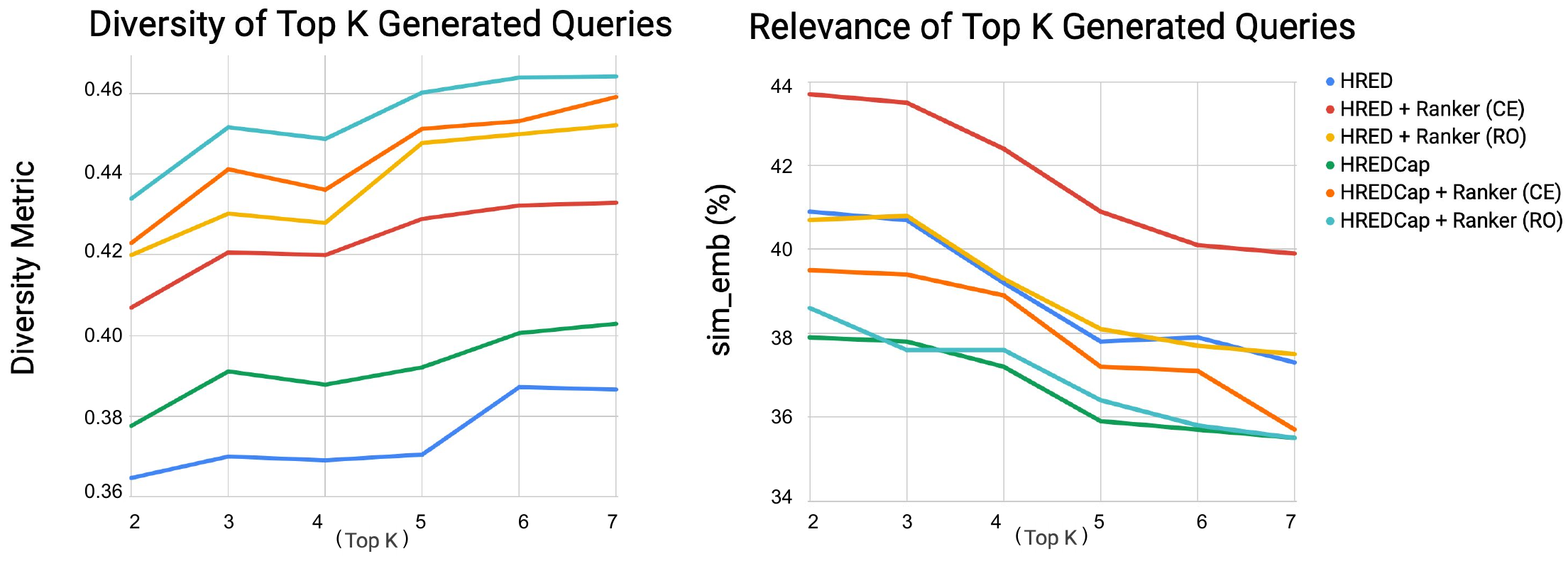}\vspace{-2mm}
\caption{\scriptsize{The trade-off between relevance (as quantified by sim$_{emb}$) and diversity. As $K$ is increased, the relevance of generated predictions drops across all models.}}
\label{fig:diversity}
\vspace{-5mm}
\end{figure*}

From Figure \ref{fig:diversity}, it can be noted that as more candidate reformulations are taken into consideration, i.e., as the beam width $K$ is increased, the average relevance of the reformulations decreases across all the models. However, the diverseness of $\mathcal{R}_{gen}$ flattens after $K=3$. This was the reason for setting the beam width to $3$ while presenting results in Table \ref{tab:mainResults}.

\vspace{-2mm}
\noindent\textbf{Descriptive Reformulations using Captions:}
The motivation for generating more descriptive reformulations is of central importance to our idea of using image captions. To this end, we analyze the generated reformulations to assess if this is indeed the case.  We start by noting (see Table \ref{tab:lengtAnalysis}) that captions corresponding to clicked images for queries in our test set contain, on average, more words than the queries. Following this, we analyze the generated reformulations by two of our multitask models -- (i) \textit{\small{HRED + Ranker (RO)}}, which guides the process of query reformulation using subsequent queries within a session, and (ii) \textit{\small{HREDCap + Ranker (RO)}}, which guides the process of query reformulation using captions corresponding to clicked images. For this entire analysis, we removed stop words \cite{nltk} 
from all the queries and captions under consideration.

As can be noted from Table \ref{tab:lengtAnalysis}, reformulations using captions tend to contain more words than reformulations without them. However, number of words in a query is only a facile proxy for its descriptiveness. Acknowledging this, we perform a secondary aggregate analysis on the number of novel words inserted into the reformulation and number of words dropped from the original query. We identify novel words as words that were not present in the original query $q_i$ but have been generated in the reformulation $\hat{q}_{\text{reform}}$, and dropped words as the words that were present in the original query but are absent from the generated reformulation. Table \ref{tab:lengtAnalysis} indicates that, on average, the model trained using captions tends to insert more novel words while reformulating the query, and at the same time drops fewer words from the query. Interestingly, models trained using subsequent queries inserts almost as many words into the reformulation as it drops from the original query. 

To analyze this further, we compute the average similarity between the novel words that were inserted and the words that were dropped,
by averaging the GloVe vector based similarity between words, across all  queries in our test set. For \textit{\small{HRED + Ranking (RO)}} this average similarity is $\mathbf{0.64}$, while for \textit{\small{HREDCap + Ranker (RO)}} it is $\mathbf{0.41}$. A higher similarity value for the former suggests that the model largely \textit{substitutes} the existing words with words having similar semantic meaning. Using captions, on the other hand, is more likely to generate novel words which bring in additional meaning.

\begin{table*}[!t]
    \centering
    \scalebox{0.68}{
    \begin{tabular}{| c | c | c |}\hline
    Avg. \# of words in queries & \multicolumn{2}{c |}{$2.31 \pm 0.92$ word(s)} \\
    Avg. \# of words in captions &  \multicolumn{2}{c |}{$5.22 \pm 2.37$ word(s)}\\\hline\hline
    \textbf{Models $\rightarrow$} & {HRED + Ranker (RO)}$\text{ }$ & $\text{ }${HREDCap + Ranker (RO)}\\\hline
    Avg. \# generated words    & $2.18 \pm 0.61 $ word(s) & $4.91 \pm 1.16$ word(s) \\
    Avg. \# novel words   & $1.04 \pm 0.13$ word(s) & $2.56 \pm 0.47$ word(s) \\
    Avg. \# dropped words & $1.14 \pm 0.15$ word(s) & $0.89 \pm 0.17$ word(s)\\\hline
    Avg. similarity b/w insertions and drops & $0.64 \pm 0.03$ & $0.41 \pm 0.04$\\\hline
    \end{tabular}
    }\vspace{0.5mm}
    \caption{\scriptsize{Analyzing the effect of using captions on length of generated query reformulations, along with influence on generating novel words while dropping the existing ones. }}
    \label{tab:lengtAnalysis}
\vspace{-10mm}
\end{table*}

\vspace{-2mm}
\subsubsection{5.4 \quad Qualitative Results:}
\vspace{-3.5mm}

In Table \ref{tab:qualResults}, we present a few examples depicting the descriptive nature of generated reformulations. The generated reformulations by \textit{\small{HRED + Ranker}} are compared against those  by \textit{\small{HREDCap + Ranker}}. We only present the top ranked reformulation among top-$K$ reformulations.
We note that using captions as target generates reformulations that are more descriptive and the process of generation results in more insertions of novel words, in comparison to using subsequent queries as targets. These qualitative observations, along with quantitative observations discussed earlier, reinforce the efficacy of using captions of clicked images for the task of query reformulation.

\begin{table*}[!h]\vspace{-5mm}
    \centering
    \scalebox{0.55}{
    \begin{tabular}{|c | c  p{3.4cm} | p{7.0cm} | p{3.4cm} | p{6.4cm} |}\hline
        {} & \multicolumn{2}{c|}{\textbf{Queries}} & \textbf{Clicked Caption} & \textit{HRED + Ranker (RO)} & \textit{HREDCap + Ranker (RO)} \\\hline
        \multirow{3}{*}{$\mathbf{Session_1}$} & $\mathbf{q_1}$ & traffic  & rush hour traffic & traffic jam & traffic \textbf{jam during rush hour}  \\
        
        & $\mathbf{q_2}$ & traffic \textbf{jam}  & traffic jams in the city, road, rush hour
 & \textbf{city} traffic jam & traffic \textbf{during} \textbf{rush hour} in \textbf{city} \\
 
        & $\mathbf{q_3}$& traffic jam pollution & blurred silhouettes of cars by steam of exhaust
 & traffic jam \textbf{cars} & \textbf{dirt} and \textbf{smoke} from \textbf{cars} in traffic jam\\\hline
 
        \multirow{3}{*}{$\mathbf{Session_2}$} & $\mathbf{q_1}$ & sleeping baby & sleeping one year old baby girl & \textbf{cute} sleeping baby & \textbf{little} baby sleeping \textbf{peacefully}  \\
        
        {} & $\mathbf{q_2}$ & sleeping baby cute  & baby boy in white sunny bedroom & sleeping baby & baby sleeping in \textbf{bed peacefully} \\
        
        {} & $\mathbf{q_3}$& white bed sleeping baby & carefree little baby sleeping with white soft toy & baby sleeping in bed & \textbf{little} baby sleeping in white bed \textbf{peacefully}\\\hline
        
        \multirow{2}{*}{$\mathbf{Session_3}$} & $\mathbf{q_1}$ & chemistry & three dimensional illustration of molecule model & chemical \textbf{reaction} & \textbf{molecules} and \textbf{structures} in chemistry  \\
        
        {} & $\mathbf{q_3}$& molecule reaction & chemical reaction between molecules & reaction molecules & molecules reacting in \textbf{chemistry}\\
        
        {} & $\mathbf{q_3}$& molecule collision & frozen moment of two particle collision & collision molecules & molecules colliding \textbf{chemistry} \textbf{reaction}\\\hline
    \end{tabular}
    }\vspace{0.5mm}
    \caption{\scriptsize{Qualitative results comparing the generated reformulation by \textit{HRED + Ranker} and \textit{HREDCap + Ranker}. The words in \textbf{bold} are novel insertions.}}
    \label{tab:qualResults}\vspace{-9mm}
\end{table*}

%
\vspace{-6mm}
\section{Conclusion}
\vspace{-4mm}

In this paper, we build upon recent advances in sequence-to-sequence models based approaches for recommending queries. The core technical component of our paper is the use of a novel supervision signal for training seq-to-seq models for query reformulation --  i.e., captions of clicked images instead of subsequent queries within a session, as well as the use of a pairwise preference based objective for the secondary ranking task. The effect of these are evaluated alongside baseline model architectures for this setting. Our extensive analysis evaluated the model and training method combinations towards being able to generate a set of descriptive, relevant and diverse reformulations.

Although the experiments were done on data from an image search engine, we believe that similar improvements can be observed if content properties from textual documents can be integrated into the seq-to-seq models. 
Future work will look into the influence of richer representations on the behavior of the ranker, and in turn on the characteristics of the reformulations. %

%
%
%
\bibliographystyle{splncs04}
\bibliography{mybibliography}
%




\end{document}